\documentclass[conference,twocolumn]{IEEEtran}
\usepackage[T1]{fontenc}
\usepackage{amsmath,bm,amsfonts,amssymb,graphicx,color,multirow,setspace,xfrac,comment}
\usepackage[a4paper, right=0.9in, left=0.9in,top=0.9in, bottom=0.9in]{geometry}

\usepackage{booktabs}
\usepackage{epstopdf}
\usepackage{mathtools}
\usepackage{balance}
\usepackage{lipsum}
\usepackage{cite}
\usepackage{citesort}
\newcommand*{\QED}{\hfill\ensuremath{\blacksquare}}

\begin{document}
\title{Channel Correlation Diversity in MU-MIMO Systems -- Analysis and Measurements \vspace{-3pt}}
\author{\IEEEauthorblockN{Harsh~Tataria\IEEEauthorrefmark{1},		
						Seun~Sangodoyin\IEEEauthorrefmark{2},  
  						Andreas~F.~Molisch\IEEEauthorrefmark{3},
  						Peter~J.~Smith\IEEEauthorrefmark{4},
  						Michail~Matthaiou\IEEEauthorrefmark{5},\\
  						Jianzhong~Zhang\IEEEauthorrefmark{6}, and
  					      Reiner.~S.~Thom\"{a}\IEEEauthorrefmark{7}
  									}
  \IEEEauthorblockA{\IEEEauthorrefmark{1}Department of Electrical and Information Technology, Lund University, Lund, Sweden}    \IEEEauthorblockA{\IEEEauthorrefmark{2}School of Electrical and Computer Engineering, Georgia Institute of Technology, Atlanta, GA, USA}
  \IEEEauthorblockA{\IEEEauthorrefmark{3}Department of Electrical Engineering, University of Southern California, Los Angeles, CA, USA}
   \IEEEauthorblockA{\IEEEauthorrefmark{4}School of Mathematics and Statistics, Victoria University of Wellington, Wellington, New Zealand}
 \IEEEauthorblockA{\IEEEauthorrefmark{5}Institute of Electronics, Communications and Information Technology, Queen's University Belfast, Belfast, UK}
  \IEEEauthorblockA{\IEEEauthorrefmark{6}Mobility and Innovation Laboratory, Samsung Research America, Richardson, TX, USA}
    \IEEEauthorblockA{\IEEEauthorrefmark{7}Institut f\"{u}r Informationstechnik, Technische Universit\"{a}t Ilmenau, Ilmenau, Germany}  
 \IEEEauthorblockA{e--mail:
 harsh.tataria@eit.lth.se,~seun.sangodoyin@gatech.edu, molisch@usc.edu,~peter.smith@vuw.ac.nz,\\m.matthaiou@qub.ac.uk,~jianzhong.z@samsung.com,~and~reiner.thomae@tu-ilmenau.de}\vspace{-20pt}}

\maketitle

\begin{abstract}
In multiuser multiple--input multiple--output (MU--MIMO) 
systems, \emph{channel correlation} is detrimental to  
system performance. 
We demonstrate that widely used, yet overly 
simplified, correlation models that generate \emph{identical}
correlation profiles for each terminal tend to severely 
underestimate the system performance. In sharp 
contrast, more physically motivated models that  
capture \emph{variations} in the power angular 
spectra across multiple terminals, generate 
\emph{diverse} correlation 
patterns. This has a significant impact on the system 
performance. Assuming correlated 
Rayleigh fading and downlink zero--forcing precoding, tight 
closed--form approximations for the average 
signal--to--noise--ratio, 
and ergodic sum spectral efficiency are derived. 
Our expressions provide clear \emph{insights} 
into the \emph{impact} of diverse correlation patterns on the above performance 
metrics. Unlike previous works, the correlation 
models are parameterized with \emph{measured} data from a 
recent 2.53 GHz urban macrocellular campaign in 
Cologne, Germany. Overall, results from this paper can 
be treated as a timely re--calibration of performance 
expectations from practical MU--MIMO systems. 
\end{abstract}
\IEEEpeerreviewmaketitle

\vspace{-1pt}
\section{Introduction}
\label{Introduction}
\let\thefootnote\relax\footnote{\copyright\hspace{1pt}
2019 IEEE. This paper was originally accepted for publication at IEEE GLOBECOM 2018, Abu Dhabi, December 2018, where it was \emph{not presented}, and thus not included in IEEE Xplore. As such, the paper is now accepted for publication in IEEE PIMRC 2019, Istanbul, September 2019, with permission from the IEEE Copyrights Office and from IEEE GLOBECOM 2018 Technical Program Committee Chair.}
It is now well understood that multiuser 
multiple--input multiple--output (MU--MIMO) 
systems with large antenna arrays at the base 
station (BS) will form an integral part of fifth--generation cellular \cite{SHAFI2,LARSSON1}. In its most general form, 
propagation between the BS and user terminals 
can be characterized by far--field multipath 
components (MPCs) arriving at the terminals from 
set of objects (a.k.a. scatterers) 
\emph{interacting} with the transmitted waveform 
\cite{GAO1,SHAFI1}. MPCs often 
arrive at multiple terminals from \emph{clusters} of 
scatterers, leading to \emph{non--uniform} power angular  
spectra as seen by the BS, resulting in \emph{correlation} of signals at the BS array elements \cite{ASPLUND1}. 
Furthermore, depending on the severity of 
scattering in a given environment, 
and the relative physical separation 
of terminals, MPCs to \emph{different} terminals arrive 
via the \emph{same} cluster of scatterers, and 
hence are correlated \cite{NGUYEN1,GAO2}. 
From single--user MIMO literature, it is known that 
channel correlation has a negative impact on the 
terminal and system spectral efficiency (see e.g., 
\cite{GAO1,NGUYEN1,GAO2,FALCONET1,TATARIA1,TATARIA3}). 

In stark contrast, a different set of investigations has 
shown that correlation can \emph{enhance} MU--MIMO performance 
\cite{NAM1,ADHIKARY1,ADHIKARY2,BJORNSON1,POUTANEN1,ISCAR1,TATARIA2,RAGHAVAN1,CHIEN1,NAM2}. 
Collectively, a critical observation from these 
studies is that the departing spread of electromagnetic energy 
from the BS can arrive at the terminals with substantially 
\emph{different} power angular spectra, leading to 
\emph{variations} in the channel statistics.
Fundamentally, such variations depend on the \emph{geometry} of scattering, as well as the \emph{inter--element spacing} at the BS. To encapsulate these variations power angular spectra, models such as one--ring correlation 
have been proposed \cite{NAM1,ADHIKARY1,ADHIKARY2,BJORNSON1,TATARIA2,NAM2}. 
Such models are characterized in terms of the mean 
direction--of--arrival (DOA) at a terminal, angular spread of departure, 
as well as the antenna spacing at the BS. 
The authors of \cite{NAM1,ADHIKARY1} utilized the one--ring model to 
group the terminals with similar correlation characteristics in order to motivate 
the joint spatial division multiplexing technique. 
The study in \cite{BJORNSON1} shows that if correlation matrices span \emph{orthogonal} subspaces, pilot contamination can vanish, 
allowing the ergodic spectral efficiency to grow without bound with increasing numbers 
of BS antennas. The net spectral 
efficiency of a MU--MIMO system was \emph{numerically} investigated 
in \cite{ISCAR1} as a function of the azimuth angular spread with pilot 
contamination. 

Despite the above efforts, it remains to be seen just 
\emph{how much} performance gain is available with correlation diversity, relative 
to the case when each terminal has an identical correlation pattern. More importantly, 
almost all of the \emph{analytical results} predicting MU--MIMO performance with 
correlation diversity, are left in terms of \emph{complex} mathematical expressions 
making it difficult 
to gain any insights into their behavior 
(see e.g., \cite{ADHIKARY1,HOYDIS1,WAGNER1}). 
To gain a fundamental understanding of MU--MIMO performance with 
and without correlation diversity, it is desirable to have an insightful and 
simple downlink performance measure. This is largely missing from the 
literature, and hence is the aim of the paper. 
With this in mind, we derive \emph{simple}, closed--form approximations 
to the downlink zero--forcing (ZF) expected signal--to--noise--ratio 
(SNR) and ergodic sum spectral efficiency. It is noteworthy that 
our recent work in \cite{TATARIA2} makes an initial attempt 
to explore the aforementioned issues with simple \emph{matched filter}
transmission, where we derive approximations for 
determining the impact of correlation diversity on MU--MIMO 
performance. Unlike \cite{TATARIA2}, here we 
consider the more challenging case of ZF transmission, and examine 
the MU--MIMO gains with diversity in the channel correlation structure. 
As more physical correlation 
models rely on the propagation channel's spatial parameters, for the most accurate 
parameterization, we extract the required parameters from a recent 2.53 GHz 
MU--MIMO \emph{measurement campaign} in Cologne, Germany. 
To the best of the our knowledge, 
studies which use \emph{measured} multipath parameters to investigate the 
diversity of correlation profiles in multiuser systems are rare.

\textbf{Contributions.} Assuming spatially correlated Rayleigh fading, our ZF 
approximations provide clear \emph{insights} into the \emph{impact} 
of correlation diversity, as well as other system parameters, 
such as the number of BS antennas, number of terminals, 
and the average operational downlink SNR. 
We provide \emph{explicit} insights into the fact that fixed correlation profiles tend to \emph{amplify} 
the expected ZF noise power, unlike diverse correlation 
profiles. We therefore argue that fixed correlation profiles can be used as a 
useful \emph{lower bound} on the resulting system performance. 
Our results disclose that the \emph{choice} of a  
particular correlation model has a direct relation to the expected ZF SNR 
and ergodic sum spectral efficiency. More physical models  
such as one--ring correlation, give superior performance over more 
simple models, such as 
the exponential and Clerckx correlation \cite{TATARIA1,CLERCKX1}. 
To parameterize the correlation models, 
we utilize \emph{measured} angular spreads and mean DOA distributions 
at 2.53 GHz from an urban macrocellular (UMa) measurement campaign. 

\textbf{Notation.} Upper and lower boldface letters represent matrices and vectors. 
The $M\times{}M$ identity matrix is denoted as $\mathbf{I}_{M}$. Transpose, 
Hermitian transpose, inverse and trace operators are denoted by $(\cdot)^{T}$, 
$\left(\cdot\right)^{H}$, $(\cdot)^{-1}$, and $\textrm{Tr}[\cdot]$, 
respectively. Moreover, $||\cdot{}||_{F}$ denotes the Frobenius 
norm. We use $\mathbf{h}\sim\mathcal{CN}\left(
\mathbf{m},\mathbf{R}\right)$ to denote a complex Gaussian distribution for $\mathbf{h}$ 
with mean $\mathbf{m}$ and covariance matrix $\mathbf{R}$. Similarly, 
$h\sim\mathcal{U}\left[a,b\right]$ is used to denote a uniform random variable for 
$h$, taking on values from $a$ to $b$. Finally, $\mathbb{E}\{\cdot\}$ 
denotes the statistical expectation.

\vspace{-10pt}
\section{System Model}
\label{SystemModel}
We consider the downlink of a single--cell MU--MIMO system operating 
in an UMa environment. The BS is located at the center of a circular cell 
with radius $R_{\textrm{c}}$, and 
is equipped with an array of $M$ transmit antennas. The BS serves $L$ 
single--antenna terminals $\left(M\geq{}L\right)$ in the same time--frequency 
resource. Channel knowledge is assumed at the BS with narrow--band transmission 
and uniform power allocation. 

\textbf{Remark 1.} \emph{At first sight, the assumption of perfect channel knowledge may 
seem rather naive. However, there are several reasons for this: 
Firstly, unlike previous studies, the central focus of the work is to devise 
a simple, yet accurate, performance metric to gain insights into the behavior 
of correlation diversity in multiuser systems. In contrast to prior studies, 
measured spatial parameters of the channel are utilized to capture the 
power angular spectra variations across multiple terminals. Under this 
heterogeneous scenario, it is extremely difficult to make 
analytical progress without perfect channel knowledge. Secondly, 
this assumption allows us to effectively separate the propagation 
effects from the system related effects, i.e., to study the influence 
of correlation diversity in isolation. 
Thirdly, it is worth noting that the results obtained from the subsequent 
analysis and evaluations can be treated as a useful upper bound on 
the performance which may be seen in practice with estimated 
propagation channels.}

The $1\times{}M$ propagation channel to terminal $\ell$ from the 
BS array is denoted by $\mathbf{h}_{\ell}$, which is 
assumed to follow a spatially correlated Rayleigh fading 
distribution, i.e., $\mathbf{h}_{\ell}\sim\mathcal{CN}
\left(\mathbf{0},\mathbf{R}_{\ell}\right)$. Unlike previously 
(see e.g., \cite{FALCONET1,TATARIA1}), 
$\mathbf{R}_{\ell}$, the $M\times{}M$ correlation matrix, is 
\emph{specific} to terminal 
$\ell$. Naturally, $\mathbf{R}_{\ell}$ will be a function of the channel's 
spatial parameters 
\cite{NAM1,ADHIKARY1,BJORNSON1,NGUYEN1}. 
For clarity, further discussion on the possible structures of 
$\mathbf{R}_{\ell}$ is deferred till   
Sec.~\ref{NumericalResults}. 
The received signal at terminal $\ell$ is
\vspace{-1pt}
\begin{equation}
\label{downlinkreceivedsignal}
y_{\ell}=\sqrt{\frac{\rho_{\textrm{t}}\hspace{1pt}
\beta_{\ell}}{\eta}}\hspace{1pt}\mathbf{h}_{\ell}\hspace{1pt}
\mathbf{g}_{\ell}\hspace{1pt}
s_{\ell}+\sum\limits_{\substack{i=1\\i\neq{}\ell}}^{L}
\sqrt{\frac{\rho_{\textrm{t}}\hspace{1pt}
\beta_{\ell}}{\eta}}\hspace{2pt}
\mathbf{h}_{\ell}\hspace{1pt}
\mathbf{g}_{i}
s_{i}+n_{\ell},
\vspace{-4pt}
\end{equation}
where $\rho_{\textrm{t}}$ is the average transmit power 
at the BS and $\beta_{\ell}$ denotes the link gain of terminal $\ell$ 
(discussed later in the text). Furthermore, $\mathbf{g}_{\ell}$ 
is the $M\times{}1$ un--normalized downlink precoding vector from the 
BS array to terminal $\ell$, obtained from $\ell$--th column of $\mathbf{G}$, 
the composite $M\times{}L$ un--normalized precoding matrix. The data 
symbol for terminal $\ell$ is denoted by $s_{\ell}$, such that 
$\mathbb{E}\{|s_{\ell}|^{2}\}=1, \forall{}\ell=1,2,\dots,L$, and $n_{\ell} 
\sim{}\mathcal{CN}\left(0,\sigma^{2}\right)$ models the additive white Gaussian 
noise at terminal $\ell$. Note that $\sigma^{2}$ is fixed for all terminals 
$1,2,\dots{},L$. Following \cite{TATARIA1,WAGNER1}, 
$\eta=\|\mathbf{G}\|^{2}_{F}/L$ is 
the precoding \emph{normalization} parameter such that 
$\mathbb{E}\{\|\mathbf{g}_{\ell}\|^{2}\}=1$, for $\ell=1,2,\dots,{}L$ (discussed 
further in the text). 
The link gain at terminal $\ell$, 
$\beta_{\ell}=A\zeta_{\ell}\left(r_{0}/r_{\ell}\right)^{\alpha}$ is composed of 
the large--scale propagation effects: Particularly, 
$A$ is the unit--less constant for geometric attenuation at a reference 
distance $r_{0}$, $r_{\ell}$ is the link distance between the BS and terminal $\ell$, 
$\alpha$ is the attenuation exponent and $\zeta_{\ell}$ models the effects of shadow 
fading which follows a lognormal distribution, i.e., $10\log_{10}\left(\zeta_{\ell}\right)
\sim{}\mathcal{N}\left(0,\sigma_{\textrm{sh}}^{2}\right)$. For clarity, 
we delay quoting values for the above parameters 
to Sec.~\ref{NumericalResults}. 
It is well known that ZF precoding \emph{nulls}  
multiuser interference (second--term of \eqref{downlinkreceivedsignal}). 
This means that the signal--to--interference--plus--noise--ratio 
translates to a SNR \cite{TATARIA1}. Note that $\mathbf{g}_{\ell}$ 
forms the $\ell$--th column of 
$\mathbf{G}=\mathbf{H}^{H}\left(\mathbf{H}\mathbf{H}^{H}\right)^{-1}$, 
where $\mathbf{H}=
[\hspace{2pt}\mathbf{h}_{1}^{T},\mathbf{h}_{2}^{T},\dots{},
\mathbf{h}_{L}^{T}\hspace{2pt}]^{T}$ is the $L\times{}M$ matrix 
containing channels for all $L$ terminals. Recognizing that 
$\mathbf{H}\mathbf{G}=\mathbf{H}\mathbf{H}^{H}\hspace{-2pt}
\left(\mathbf{H}\mathbf{H}^{H}\right)^{\hspace{-2pt}-1}\hspace{-4pt}=
\mathbf{I}_{L}$, the ZF SNR at terminal $\ell$ is 
\vspace{-2pt}
\begin{equation}
\label{terminallsnr}
\textrm{SNR}^{\textrm{ZF}}_{\ell}=
\frac{\rho_{\textrm{t}}\hspace{1pt}\beta_{\ell}}
{\sigma^{2}\eta}=\frac{\rho_{\textrm{t}}\beta_{\ell}}
{\sigma^{2}\left\{
\frac{1}{L}\left\{\textrm{Tr}\left[\left(\mathbf{H}
\mathbf{H}^{H}\right)^{-1}\right]\right\}\right\}}, 
\end{equation}
since $\eta=\|\mathbf{G}\|^{2}_{F}/L=\textrm{Tr}
\hspace{2pt}[\hspace{1pt}
(\hspace{1pt}\mathbf{H}\mathbf{H}^{H})^{-1}
\hspace{2pt}]/L$. The ZF SNR in \eqref{terminallsnr}
can be used to estimate the ergodic sum spectral efficiency 
(in bits/sec/Hz) for all $L$ terminals. This is given by  
\vspace{-11pt}
\begin{equation}
\label{ergodicsumse}
\textrm{R}_{\textrm{ZF}}=\mathbb{E}\left\{
\sum\limits_{\ell=1}^{L}
\log_{2}\left(\hspace{1pt}1\hspace{-1pt}+\hspace{1pt}
\textrm{SNR}_{\ell}^{\textrm{ZF}}
\hspace{1pt}\right)\right\}, 
\vspace{-2pt}
\end{equation}
where the expectation is 
over \emph{small--scale} fading. 
Below, closed--form 
approximations of \eqref{terminallsnr} and 
\eqref{ergodicsumse} are derived.

\vspace{-4pt}
\section{Analytical Results and Implications}
\label{AnalyticalResultsandImplications}
\vspace{-2pt}
\subsection{Expected SNR and Ergodic Sum Spectral Efficiency}
\label{ExpectedZFSNRandErgodicSumSpectralEfficiency} 
\vspace{-1pt}
\textbf{Remark 2.} 
\emph{Finding exact moments of the ZF SNR in \eqref{terminallsnr}
is an extremely challenging task, since the matrix trace in its denominator is a random function of the inverse, of a complex non--standard semi--correlated central Wishart 
distribution formed by} $\mathbf{H}\mathbf{H}^{H}$. 
\emph{Moreover,} $\mathbf{H}$ \emph{has a fully heterogeneous structure, since it contains} $L$ \emph{different 
correlation patterns and link gains. Due to these reasons, we approximate the inverse in \eqref{terminallsnr} with a 
finite order Neumann series expansion \cite{ZHU1,KAMMOUN1}.} 
To do this, we separate $\mathbf{H}
\mathbf{H}^{H}$ into its expected diagonal 
components and correction terms. That is, 
$\mathbf{H}\mathbf{H}^{H}=M\hspace{1pt}
\mathbf{I}_{L}\hspace{1pt}+\hspace{1pt}
\mathbf{\Delta}$,
where $\mathbf{\Delta}=\mathbf{H}\mathbf{H}^{H}-
M\hspace{1pt}\mathbf{I}_{L}$ and $\mathbb{E}\left\{
\mathbf{\Delta}\right\}=\mathbf{0}$. Then, with an order $N$ 
Neumann series, we can write 
$(\mathbf{H}\mathbf{H}^{H})^{-1}$ as 
\vspace{-4pt}
\begin{equation}
\label{nsanalysis2}
\left(\mathbf{H}\mathbf{H}^{H}\right)^{\hspace{-2pt}
-1}\hspace{-2pt}
\approx\frac{1}{M}\sum\limits_{n=0}^{N}\left(
-1\right)^{n}\left(\frac{\mathbf{\Delta}}{M}\right)^{n}. 
\vspace{-1pt}
\end{equation}
Substituting the definition of $\mathbf{\Delta}$, 
and simplifying yields 
\begin{align}
\vspace{-7pt}
\nonumber
\left(\mathbf{H}\mathbf{H}^{H}\right)^{\hspace{-3pt}-1}
\hspace{-3pt}
\approx\hspace{1pt}&\frac{1}{M}\hspace{-2pt}
\sum\limits_{n=0}^{N}\frac{\left(-1\right)^{n}}
{\left(M\right)^{n}}\hspace{-3pt}
\sum\limits_{q=0}^{n}\hspace{-1pt}{n\choose{}q}\hspace{-2pt}
\left(\mathbf{H}\mathbf{H}^{H}\right)^{q}
\hspace{-2pt}
\left(-M\right)^{n-q}\\
\label{nsanalysis4x}=&\frac{1}{M}\sum\limits_{n=0}^{N}
\sum\limits_{q=0}^{n}{n\choose{}q}
\frac{\left(-1\right)^{q}}{\left(M\right)^{q}}
\left(\mathbf{H}\mathbf{H}^{H}\right)^{q}. \\[-18pt]
&\nonumber
\end{align}
Substituting the above into the denominator of 
\eqref{terminallsnr} allows us to write the ZF SNR 
for terminal $\ell$ as \eqref{nsanalysis4} (on top of 
the following page for reasons of space). 
\begin{figure*}
\begin{equation}
\label{nsanalysis4}
\textrm{SNR}_{\ell}^{\textrm{ZF}}\approx
\frac{\rho_{\textrm{t}}\hspace{1pt}\beta_{\ell}}{\sigma^{2}
\hspace{2pt}\left\{\frac{1}{L}\hspace{-1pt}
\left\{\textrm{Tr}\hspace{-1pt}\left[\frac{1}{M}
\sum\nolimits_{n=0}^{N}\sum
\nolimits_{q=0}^{n}\hspace{-2pt}
{n\choose{}q}\frac{\left(-1\right)^{q}}
{\left(M\right)^{q}}\left(\mathbf{H}\mathbf{H}^{H}
\right)^{q}\right]\right\}\right\}}. 
\vspace{3pt}
\end{equation}
\hrulefill
\vspace{-3pt}
\end{figure*}

\textbf{Remark 3.} \emph{In what follows, we evaluate the expected 
value of \eqref{nsanalysis4}. The expectation is extremely 
cumbersome to perform, since it needs to be taken over the 
myriad of small--scale fading in} $\mathbf{HH}^{H},$
\emph{a term on the denominator of \eqref{nsanalysis4}. 
To overcome this 
difficulty, we employ the univariate special case of the 
commonly used first--order Laplace approximation
\cite{TATARIA1,ZHANG1,TATARIA3}, allowing us to express 
\eqref{nsanalysis4} as \eqref{nsanalysis5} (shown 
on top of the following page for the reasons of space).}
\begin{figure*}[!t]
\vspace{-5pt}
\begin{equation}
\label{nsanalysis5}
\mathbb{E}\left\{\textrm{SNR}_{\ell}^{\textrm{ZF}}
\right\}\approx
\frac{\rho_{\textrm{t}}\hspace{1pt}\beta_{\ell}}
{\sigma^{2}\left\{\frac{1}{L}
\left\{\frac{1}{M}
\sum\nolimits_{n=0}^{N}
\sum\nolimits_{q=0}^{n}{n\choose{}q}
\frac{\left(-1\right)^{q}}
{\left(M\right)^{q}}\hspace{2pt}\mathbb{E} 
\hspace{2pt}\Big\{\textrm{Tr}\left[
\left(\mathbf{H}\mathbf{H}^{H}\right)^{q}\right]\Big\}
\right\}\right\}}.
\end{equation}
\vspace{2pt}
\hrulefill
\vspace{-6pt}
\end{figure*}
\emph{The approximation in \eqref{nsanalysis5} is a first--order 
Laplace expansion and is of the form $\mathbb{E}\{\gamma/X\}
\approx\gamma/\mathbb{E}\{X\}$, where $\gamma$ is a scalar value. 
As shown in \cite{TATARIA1,ZHANG1,TATARIA3}, the accuracy of such 
approximations relies on $X$ having a small variance
relative to its mean. 
This can be seen by applying a multivariate Taylor series expansion to 
$\gamma/X$ around $\gamma/\mathbb{E}\{X\}$. The 
terms in \eqref{nsanalysis5} are well 
suited to this approximation, 
especially when $M$ and $L$ start to grow, since the implicit 
averaging in the denominator gives rise to the variance reduction 
required \cite{TATARIA1}. In the following proposition, 
with a two--term Neumann series 
(i.e., $N=2$), a closed--form solution to \eqref{nsanalysis5} is 
presented for heterogeneous channels.}

\textbf{Proposition 1.} \emph{When} 
$\mathbf{h}_{\ell}\sim\mathcal{CN}
\left(\mathbf{0},\mathbf{R}_{\ell}\right)$, \emph{where} 
$\mathbf{R}_{\ell}$ 
\emph{is the correlation matrix specific to terminal $\ell$, the expected 
ZF SNR for the $\ell$--th terminal can be approximated as}
\vspace{-2pt}
\begin{equation}
\label{nsanalysis6}
\vspace{-2pt}
\mathbb{E}\left\{\textrm{SNR}_{\ell}^{\textrm{ZF}}\right\}
\approx\frac{\rho_{\textrm{t}}\hspace{1pt}
\beta_{\ell}M^{3}}{\sigma^{2}\left\{L\left[
M^{2}+L\left(\textrm{Tr}\left[
\bar{\mathbf{R}}^{2}\right]
\right)\right]\right\}}, 
\end{equation}
\emph{where} 
$\bar{\mathbf{R}}=\frac{\sum\nolimits_{\ell=1}^{L}
\hspace{-2pt}\mathbf{R}_{\ell}}{L}$, \emph{and is the average correlation matrix of all terminals in the system.}
\vspace{2pt} 

\emph{Proof:} From \eqref{nsanalysis4x}, when $N=2$, one 
can write 
\vspace{-2pt}
\begin{equation}
\label{nsanalysis7}
\hspace{-2pt}
\left(\mathbf{H}\mathbf{H}^{H}\right)^{\hspace{-2pt}-1}
\hspace{-3pt}\approx\hspace{-2pt}
\frac{1}{M}\left[3\hspace{1pt}\mathbf{I}_{L}
\hspace{-2pt}-\hspace{-2pt}\frac{3}{M}\mathbf{H}\hspace{-1pt}
\mathbf{H}^{H}\hspace{-2pt}+\hspace{-3pt}
\frac{1}{M^{2}}\hspace{-1pt}\left(
\mathbf{H}\hspace{-1pt}\mathbf{H}^{H}\right)^{2}\right]. 
\end{equation}
Taking the matrix trace of \eqref{nsanalysis7} yields 
\vspace{-2pt}
\begin{align}
\nonumber
\textrm{Tr}\left[\left(\hspace{1pt}
\mathbf{H}\mathbf{H}^{H}
\right)^{\hspace{-1pt}-1}\hspace{1pt}\right]\approx
\hspace{2pt}\frac{1}{M}&
\left\{\hspace{2pt}3L-\frac{3}{M}\textrm{Tr}\left[
\mathbf{H}\mathbf{H}^{H}\right]\right.\\[-3pt]
\label{nsanalysis8}
&+\left.\frac{1}{M^{2}}
\textrm{Tr}\left[\left(\mathbf{H}\mathbf{H}^{H}
\right)^{2}\hspace{1pt}
\right]\hspace{2pt}\right\}. \\[-20pt]
&\nonumber
\end{align}
Performing the expectation of \eqref{nsanalysis8} 
results in taking the expectation of the individual 
terms on the right--hand side (RHS) of \eqref{nsanalysis8}. 
As $\mathbb{E}\{\textrm{Tr}\hspace{1pt}[\hspace{1pt}
\mathbf{HH}^{H}]\hspace{1pt}\}=M\hspace{-1pt}L$, 
the first two--terms on the RHS of \eqref{nsanalysis8} 
result in a cancellation, allowing us to focus on the 
expectation of the final term of \eqref{nsanalysis8}. 
By definition, the final term on the RHS of \eqref{nsanalysis8} is given by 
\vspace{-5pt}
\begin{equation}
\label{nsanalysis9}
\textrm{Tr}\left[\left(
\mathbf{HH}^{H}\right)^{\hspace{-1pt}2}\hspace{1pt}
\right]=
\sum\limits_{\ell=1}^{L}\sum\limits_{j=1}^{L}
\mathbf{h}_{\ell}\hspace{1pt}
\mathbf{h}_{j}^{H}
\mathbf{h}_{j}\hspace{1pt}
\mathbf{h}_{\ell}^{H}. 
\vspace{-4pt}
\end{equation}
Taking the expectation of \eqref{nsanalysis9} 
over \emph{small--scale} fading yields 
\vspace{-5pt}
\begin{align}
\nonumber
&\mathbb{E}\left\{\hspace{-1pt}
\textrm{Tr}\hspace{-1pt}\left[
\left(\mathbf{HH}^{H}\hspace{-1pt}\right)^{\hspace{-1pt}2}
\hspace{1pt}\right]\right\}\hspace{-1pt}=\hspace{-1pt}
\mathbb{E}\left\{\sum\limits_{\ell=1}^{L}
\left(\mathbf{h}_{\ell}\hspace{1pt}\mathbf{h}_{\ell}^{H}\right)^2\hspace{-3pt}\right.\\
&\hspace{90pt}+\Bigg.\sum\limits_{\ell=1}^{L}
\sum\limits_{\substack{j=1\\j\neq{}\ell}}^{L}\hspace{2pt}
\mathbf{h}_{\ell}\hspace{1pt}\mathbf{h}_{j}^{H}\mathbf{h}_{j}\mathbf{h}_{\ell}^{H}\hspace{-2pt}
\Bigg\}\\[-9pt]
&\hspace{-5pt}=\hspace{-2pt}
\mathbb{E}\left\{\hspace{-1pt}
\sum\limits_{\ell=1}^{L}
\hspace{-1pt}\left\{\hspace{-1pt}
M^{2}\hspace{-2pt}+\hspace{-2pt}
\textrm{Tr}\left[\mathbf{R}_{\ell}^{2}\right]\hspace{1pt}\right\}
\hspace{-2pt}+\hspace{-3pt}\sum\limits_{\ell=1}^{L}\sum
\limits_{\substack{j=1\\j\neq{}\ell}}^{L}\hspace{1pt}
\textrm{Tr}\left[\mathbf{R}_{\ell}\mathbf{R}_{j}\right]
\right\}\hspace{-3pt}.\\[-19pt]
&\nonumber
\end{align}
Further simplifying the above expression allows us to state 
\vspace{-2pt}
\begin{align}
\nonumber
\hspace{-2pt}\mathbb{E}\left\{\textrm{Tr}\left[
\left(\mathbf{HH}^{H}\right)^{2}\right]\right\}
=&LM^{2}+\textrm{Tr}
\left[\hspace{2pt}\sum\limits_{\ell=1}^{L}\mathbf{R}_{\ell}
\sum\limits_{j=1}^{L}\mathbf{R}_{j}\right]\\[4pt]
\label{nsanalysis10}
=&\hspace{2pt}L\left\{M^{2}+L\left(
\hspace{1pt}
\textrm{Tr}\left[
\bar{\mathbf{R}}^{2}
\right]\right)\right\}.\\[-16pt]
&\nonumber
\end{align}
Inserting \eqref{nsanalysis10} 
into the mean of 
\eqref{nsanalysis8}, and simplifying gives 
\begin{equation}
\label{nsanalysis11}
\mathbb{E}\left\{\textrm{Tr}\left[\left(
\mathbf{HH}^{H}\right)^{-1}\right]
\hspace{1pt}\right\}\hspace{-2pt}\approx
\hspace{-2pt}\frac{L}{M^{3}}\left\{M^{2}
\hspace{-2pt}+\hspace{-2pt}
L\left(\hspace{1pt}
\textrm{Tr}\left[
\bar{\mathbf{R}}^{2}
\right]\right)\right\}.  
\vspace{-2pt}
\end{equation}
The result in \eqref{nsanalysis11} can now be 
substituted into the denominator 
of \eqref{nsanalysis5} with some routine 
simplifications to obtain \eqref{nsanalysis6}. \QED
\vspace{-3pt}
\subsection{Implications of Proposition 1}
\label{ImplicationsofProposition1}
\vspace{-1pt}
To the best of the authors' knowledge, 
the result in \eqref{nsanalysis6} is the first, \emph{simple}, 
closed--form approximation with ZF precoding and 
correlation diversity. The structure 
of \eqref{nsanalysis6} readily demonstrates the 
\emph{impact} of correlation diversity via the term 
$\textrm{Tr}[\bar{\mathbf{R}}^{2}]$, which 
influences the expected noise power of the 
desired terminal. Fixing all system parameters, one 
can observe that 
the expected ZF noise power is \emph{maximized} 
when $\textrm{Tr}\left[\bar{\mathbf{R}}^{2}\right]$ is 
maximized. By definition, 
$\textrm{Tr}[\bar{\mathbf{R}}^{2}]=
M+2\sum\nolimits_{i=1}^{M-1}
\sum\nolimits_{j=i+1}^{M}|\bar{r}_{i,j}|^{2}$,  
where $\bar{r}_{i,j}$ denotes the $(i,j)$--th 
element of $\bar{\mathbf{R}}$, such that 
$|\bar{r}_{i,j}|^{2}=
|\frac{1}{L}\sum\nolimits_{\ell=1}^{L}
(\mathbf{R}_{\ell})_{i,j}|^{2}$.
Maximizing $|\bar{r}_{i,j}|^{2}$ requires 
\emph{alignment} of all the terms in the 
modulus. Specifically, the \emph{phases} 
of \emph{all} entries in $\mathbf{R}_{\ell}$ 
need to align in the $(i,j)$--th position, 
$\forall{}\ell=1,2,\dots{},L$. 
While such a scenario is \emph{generally unlikely} 
to occur in practice, we identify one possible situation when this 
may take place: Precisely, when each terminal's 
correlation matrix is \emph{identical} (the case for homogeneous 
channels), \emph{all} phases will be aligned across \emph{all} 
terminals in the $(i,j)$--th position. Though the above 
condition \emph{does not} require the amplitudes of each terminal's 
correlation matrix to be \emph{equal}, in the case of identical 
correlation matrices, the amplitudes will naturally also be 
equal across all terminals. Hence, identical correlation 
matrices result in the \emph{lowest} ZF SNRs, serving 
as a useful \emph{lower bound} 
for the performance of correlated multiuser channels. 
On the other hand, when each terminal experiences the \emph{same} 
angular spread, yet a \emph{different} mean angle, 
$\textrm{Tr}[\bar{\mathbf{R}}^{2}]$ yields a smaller value, 
since diversity is induced via the differences in 
the mean DOAs. 
On a similar note, with \emph{variations} in both the angular 
spread \emph{and} mean DOAs, 
\emph{maximum} diversity kicks in, where  
$\textrm{Tr}[\bar{\mathbf{R}}^{2}]$ 
tends to be even smaller, 
leading to a higher ZF SNR. 

In addition to the above, holding all other propagation and 
system parameters constant, 
increasing the number of BS antennas, $M$, 
causes a linear increase in the expected SNR (due to its 
numerator containing $M^{3}$ and the 
denominator containing a $M^{2}$).  
Meanwhile, increasing the number of 
user terminals, $L$, while keeping other parameters 
fixed leads to a quadratic increase in the expected 
noise power, degrading the ZF SNR. 
The above insights are difficult to obtain 
from more complex solutions derived 
in the literature (see e.g., 
\cite{ADHIKARY1,HOYDIS1,WAGNER1}), 
which require a linked set of 
equations, even in the large system regime.
In contrast, our analysis poses no such constraints 
and is \emph{general} to the operational system 
dimension, and channel correlation structure. 
Note that \eqref{nsanalysis6} can be used 
to approximate the ergodic sum spectral 
efficiency given by
\vspace{-6pt} 
\begin{equation}
\label{ergodicsumspectraleffapprox}
\mathbb{E}\left\{\textrm{R}_{\textrm{ZF}}\right\}\approx
\sum\limits_{\ell=1}^{L}\log_{2}\left(1+\mathbb{E}\left\{
\textrm{SNR}_{\ell}^{\textrm{ZF}}\right\}\right). 
\vspace{-4pt}
\end{equation}
The accuracy of the derived results in 
\eqref{nsanalysis6} and 
\eqref{ergodicsumspectraleffapprox} with their 
simulated counterparts is presented in 
Sec.~\ref{NumericalResults}. 

\textbf{Remark 4.} 
\emph{In addition to the 
results presented in Sec.~\ref{NumericalResults}, 
we note that \eqref{nsanalysis6} and 
\eqref{ergodicsumspectraleffapprox} were explicitly tested with sample simulations based on 
scattering cluster circles in the propagation channel 
(one circle per--terminal) with varying center distances 
from each other. The one--ring correlation 
model described in 
Sec.~\ref{ChannelCorrelationModels} was 
used for the aforementioned evaluations, where 
agreement within 1 dB was found from the 
results derived in \eqref{nsanalysis6} and 
\eqref{ergodicsumspectraleffapprox} relative to the 
simulated cases.} 

\vspace{-4pt}
\section{Channel Measurements}
\label{ChannelMeasurements}
\begin{figure*}
    \centering
    \includegraphics[width=12cm]{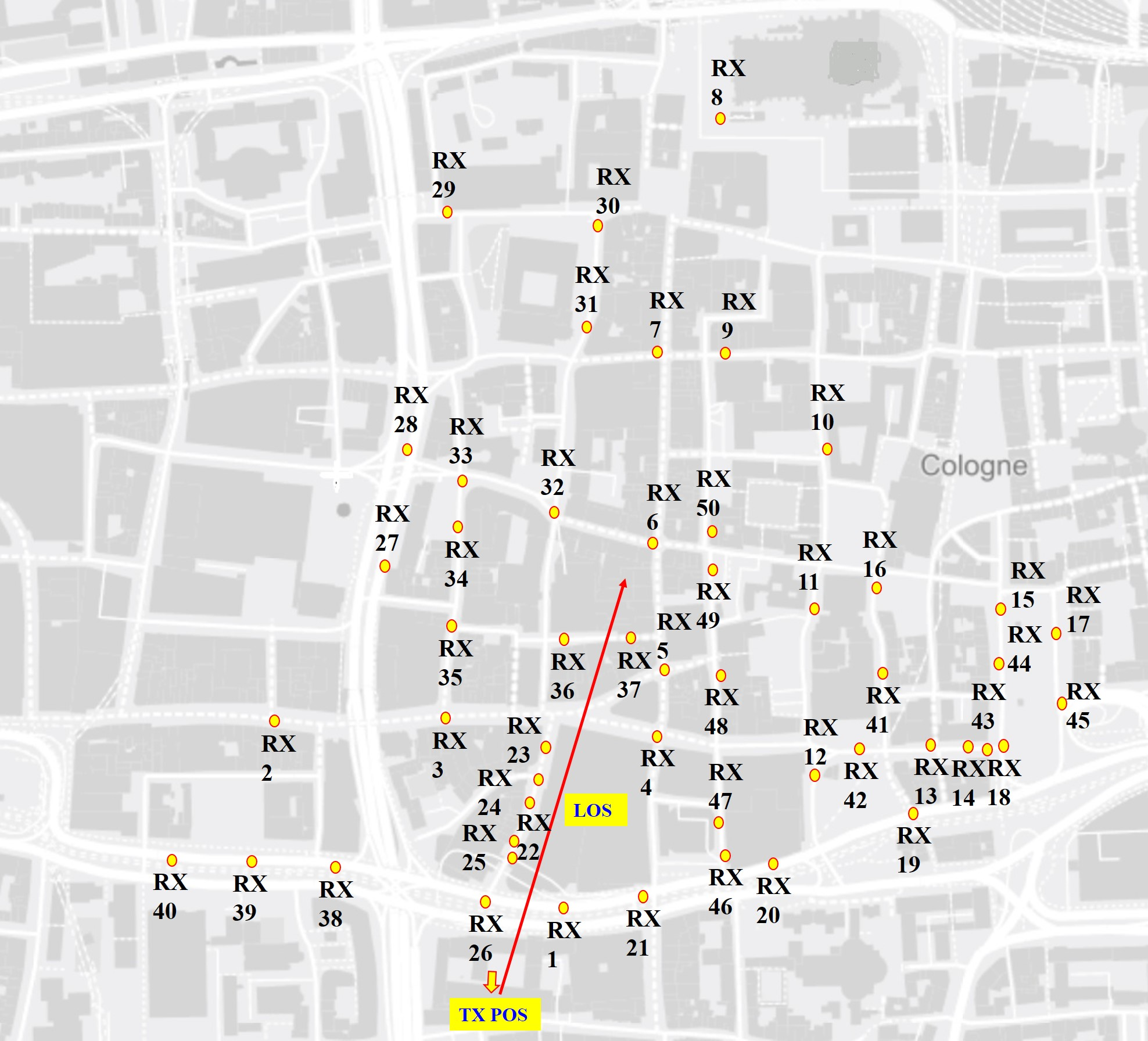}
    \caption{Birds eye view of the massive MIMO 
    channel measurement campaign across 50 different RX  positions.}
    \label{RXpositions}
\end{figure*}

\vspace{-2pt}
We will evaluate our analytical results on MU--MIMO channel measurements in an urban macrocellular environment. The measurements were performed in the city of Cologne in Germany. The BS acted as TX, and was placed on a high--rise building, and was thus located 30 m above ground (see e.g., \cite{TATARIA2,SANGODOYIN1,SANGODOYIN2} for a picture). The buildings in the area of interest were (with the exception of the high rise on which the BS was located, and the Cologne cathedral) of approximately similar height, ranging from 4--8 floors.  The terminal, acting as RX, was on the rooftop of a car, roughly 2.5 m above ground (see e.g., \cite{TATARIA2,SANGODOYIN1,SANGODOYIN2} for a picture). The RX was placed at 50 distinct locations throughout the cell, distributed across 800 m from the TX. The distribution of RX positions throughout the measurement campaign is depicted in Fig.~\ref{RXpositions}. Note that line-of-sight can be ``seen" at multiple positions, as shown in the figure. 
Since the impact of moving scatterers was negligible, the channels for the different RX positions can be used to emulate a multiuser propagation measurement campaign with service across 50 terminals. The channel sounding was performed with a MEDAV RUSK channel sounder for wideband MIMO channels. The sounder operated with a center frequency of 2.53 GHz, and a bandwidth of 20 MHz; since the operator (Deutsche Telekom) owning this band switched off the surrounding BSs for the duration of the measurements, no interference was present. The arrays used at both link ends were cylindrical, with 8 vertical and 60 horizontal antenna elements at the TX, and 2 vertical and 8 horizontal elements at the RX; all elements were dual polarized. Due to the cylindrical structures, MPCs from all directions (all azimuths and elevations) could be received and measured. The sounding principles are slightly different at TX and RX: the RX uses the switched antenna array principle \cite{THOMA1,SANGODOYIN1}, such that a single antenna element is measured at a time; an electronic switch is used to quickly cycle through all RX antenna elements and connect them sequentially to the {\em single} existing radio--frequency downconversion chain. The TX, on the other hand, uses a combination of this switched sounding (for the vertical and polarization domains) with a virtual array in azimuth, created by mechanical rotation of the physical (vertical) array. Measurement of a MIMO snapshot (all combinations of TX and RX antennas) took about 10 minutes; this long duration was acceptable because the high terminal position mostly eliminated the impact of moving scatterers and the channel was thus essentially static. More details about the measurement setup and the environment can be found in \cite{SANGODOYIN2}.  

\begin{figure}[!t] 
\vspace{-4pt}
\centering
\hspace{-11pt}
\includegraphics[width=8.2cm]{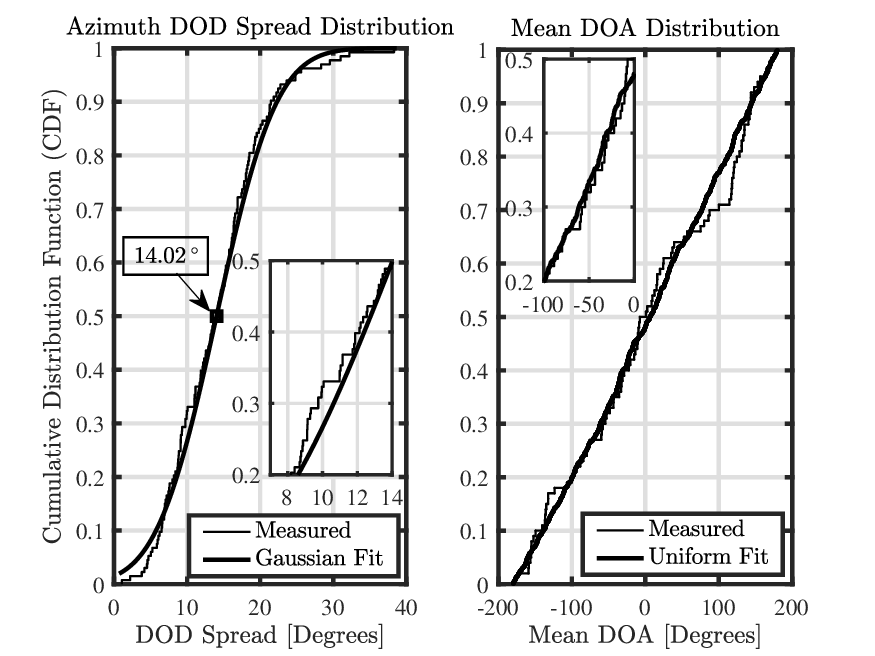}
\caption{Measured and fitted azimuth DOD spread and mean 
DOA CDFs at 2.53 GHz in an UMa environment in Cologne, as in
\cite{TATARIA2}.}
\label{angularSpreadMeanAoA}
\vspace{-3pt}
\end{figure}

The channel sounder measures the channel transfer function matrix.   
$\mathbf{H}(f,t,r)$; multiple snapshots of that matrix are averaged to improve the SNR.  Here $f$ is the index of the subcarrier at which $\mathbf{H}$ is evaluated, 
and $t$ and $r$ denote the indices of the TX and RX antenna elements, respectively. For simplicity, we do not explicitly denote polarization henceforth. From the 
transfer function matrix, we extract the parameters of the MPCs via RiMAX 
\cite{RICHTER1}, a high resolution parameter estimation algorithm that can be interpreted as an iterative maximum--likelihood estimator. 
RiMAX provides a double--directional channel description \cite{STEINBAUER1}, i.e., an \emph{antenna independent}
characterization of the channel. In particular, it provides the amplitude, delays, DOAs, and direction--of--departures (DODs) of all 
MPCs. We refer the reader to \cite{SANGODOYIN1,SANGODOYIN2} for a more detailed description of the parameter extraction routines. From the double--directional channel characterization, we can extract the \emph{root mean square} azimuth angular spread of 
departure, as well as the mean DOA distributions across all 50 terminal positions. The obtained distributions and functional fits are depicted in fig.~\ref{angularSpreadMeanAoA}. Rather interestingly, we can observe variability in the azimuth angular spread of departure and mean DOA, which can 
be modeled as $\mathcal{N}(14.02^{\circ},(6.45^{2})^{\circ})$ and $\mathcal{U}[-180^{\circ}, 180^{\circ}]$, primarily reflecting the multipath characteristics at the TX and RX positions, respectively. From these we can obtain the parameterized 
correlation structures considered in the paper 
(see Sec.~\ref{ChannelCorrelationModels}).


\vspace{-2pt}
\section{Numerical Results}
\label{NumericalResults}
Unless otherwise specified, the parameters described 
below are utilized for all numerical results, and are 
obtained from \cite{3GPP1}. The cell radius, 
$R_{\textrm{c}}=500$ m was 
chosen with a reference distance $r_{0}=50$ m, 
such that terminals are randomly located outside 
$r_{0}$ and inside $R_{c}$, following 
$\mathcal{U}[-180^{\circ},180^{\circ}]$. Uniform 
power control is assumed so that the average 
transmit power is independent of distance.  
The UMa attenuation exponent of $\alpha=3.67$ 
was chosen. Furthermore, it is assumed that 
$\sigma^{2}=1$, and hence the average downlink 
SNR is \emph{equivalent} to the average downlink 
transmit power, $\rho_{\textrm{t}}/\sigma^{2}
=\rho_{\textrm{t}}$. The unit--less constant for 
geometric attenuation, $A$, is chosen such that 
the fifth percentile of the instantaneous SNR with 
ZF precoding at terminal $\ell$ is 0 dB, 
when $\rho_{\textrm{t}}=0$ dB with the baseline system 
parameters of $M=64$ and $L=6$. Note that 
the exponential correlation model (described 
later, with the correlation coefficient 
$\xi=0.9$ was used to obtain $A$. The 
shadowing standard deviation, 
$\sigma_{\textrm{sh}}=6$ dB. For all 
numerical results, $10^{5}$ Monte--Carlo 
realizations were generated with an inter--element 
spacing, $d=0.5\lambda$ at the BS, where applicable 
with $\lambda$ denoting the 
wavelength at the desired frequency. 
\vspace{-1pt}
\subsection{Channel Correlation Models}
\label{ChannelCorrelationModels}
As a baseline case, we assign a fixed correlation profile 
to each terminal with the widely used exponential model. Here,  
the $(i,j)$--th element of $\mathbf{R}_{\ell}$ is expressed as 
$\left[\mathbf{R}_{\ell}\right]_{i,j}=\xi^{|i-j|}$, 
for any $i,j$ in $1,2,\dots{},M$ with $0\leq\xi\leq{}1$ 
\cite{TATARIA1}.  
With correlation diversity, we employ two models, 
namely \emph{Clerckx} \cite{CLERCKX1}, and \emph{one--ring}  
\cite{ADHIKARY1,BJORNSON1} 
correlation. For the Clerckx correlation model, 
$\left[\mathbf{R}_{\ell}\right]_{i,j}=\xi_{\ell}^{|i-j|}$, where 
$\xi_{\ell}=|\xi|e^{j\Delta_{\ell}}$. 
Here, $|\xi|=\xi$, as in the exponential model, and is the \emph{same} for 
each terminal. However, a \emph{terminal specific phase}, $\Delta_{\ell}$, is 
assumed to be $\mathcal{U}[-180^{\circ},180^{\circ}]$. This is used to 
differentiate the terminal locations. In each result, the range of $\Delta_{\ell}$ 
is specified. The one--ring model 
for terminal $\ell$ states $\left[\mathbf{R}_{\ell}\right]_{i,j}=\frac{1}{2\Delta_{\ell}}
 \int\nolimits_{-\Delta_{\ell}+\phi_{0}^{\ell}}^{\Delta_{\ell}+\phi_{0}^{\ell}}
 e^{-j2\pi{}d\left(i,j\right)\sin{}\left(\phi_{\ell}\right)}d\phi_{\ell}$, 
where $\Delta_{\ell}$ denotes the azimuth angular spread for terminal $\ell$, 
$\phi_{0}^{\ell}$ denotes the mean DOA, and 
$\phi_{\ell}$ is the actual DOA, uniformly distributed within the angular 
spread around the mean DOA. 
Furthermore,
$d\left(i,j\right)$ captures the \emph{normalized} antenna spacing 
between the $i$--th and $j$--th elements. 
The precise values of $\Delta_{\ell}$ for the one--ring model 
are specified in each subsequent result. 
\vspace{-1pt} 
\subsection{Impact of Correlation Diversity}
\label{ImpactofVariableCorrelationStructures}
\vspace{2pt}
\begin{figure}[!t] 
\vspace{-5pt}
\hspace{-9pt}
\centering
\includegraphics[width=8.23cm]{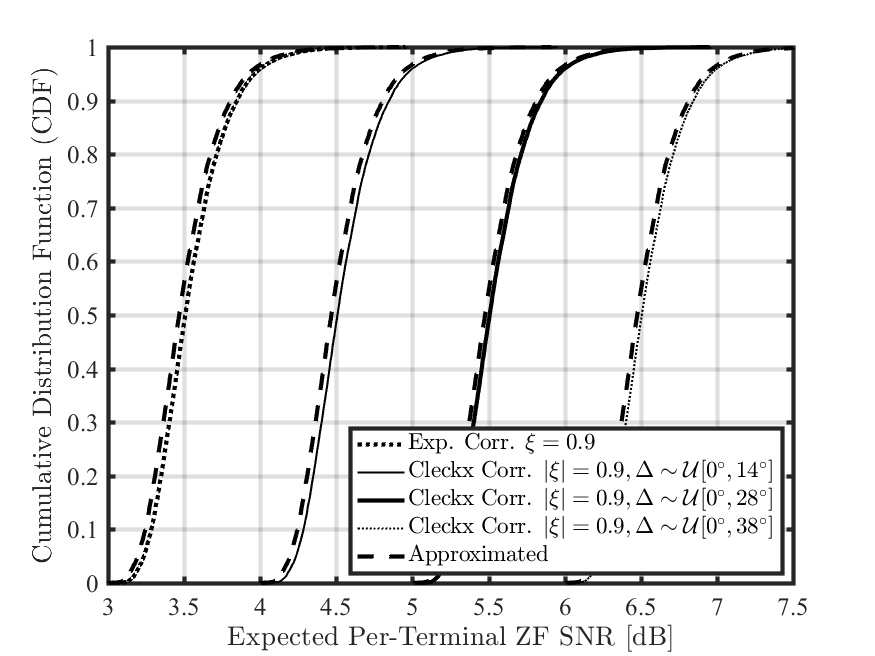}
\vspace{-10pt}
\caption{CDFs of expected ZF SNR with 
$M=64, L=6$, and $\rho_{\textrm{t}}=5$ dB.}
\label{expSNRexpClerckxComparison}
\vspace{-5pt}
\end{figure}
Figure~\ref{expSNRexpClerckxComparison} shows the CDFs 
of the expected ZF SNR with $M=64$, $L=6$, 
and $\rho_{\textrm{t}}=5$ dB. Each curve is obtained by 
averaging over the \emph{small--scale fading}, with the 
CDFs capturing the randomness from the link gains. 
\emph{Two} insights can be drawn: (1) 
With correlation diversity from the Clerckx 
model, the \emph{larger} the spread of the random phases in 
$\Delta$'s, the \emph{higher} the expected ZF SNR. 
Irrespective of the correlation magnitude 
being as high as $\xi=0.9$, increasing the spread of $\Delta$ to 
$\mathcal{U}\left[0,14^{\circ}\right]$, $\mathcal{U}\left[0,28^{\circ}\right]$, 
and $\mathcal{U}\left[0,38^{\circ}\right]$ yields a 0.93, 2, and 3 dB gain 
in the expected ZF SNR, relative to fixed correlation (exponential model) patterns. 
This difference is due to the increasing the spread of 
$\Delta$, consequently increasing the amount of \emph{path selectivity} 
across multiple channels, yielding higher composite channel rank. 
The result demonstrates the sensitivity of MU--MIMO to changes in the 
phase of the correlation patterns. Secondly, our proposed approximations 
agree well with the simulated cases, for all values of $\Delta$. 

\begin{figure}[!t]
\vspace{-10pt}
\hspace{-15pt}
\centering
\includegraphics[width=8.29cm]{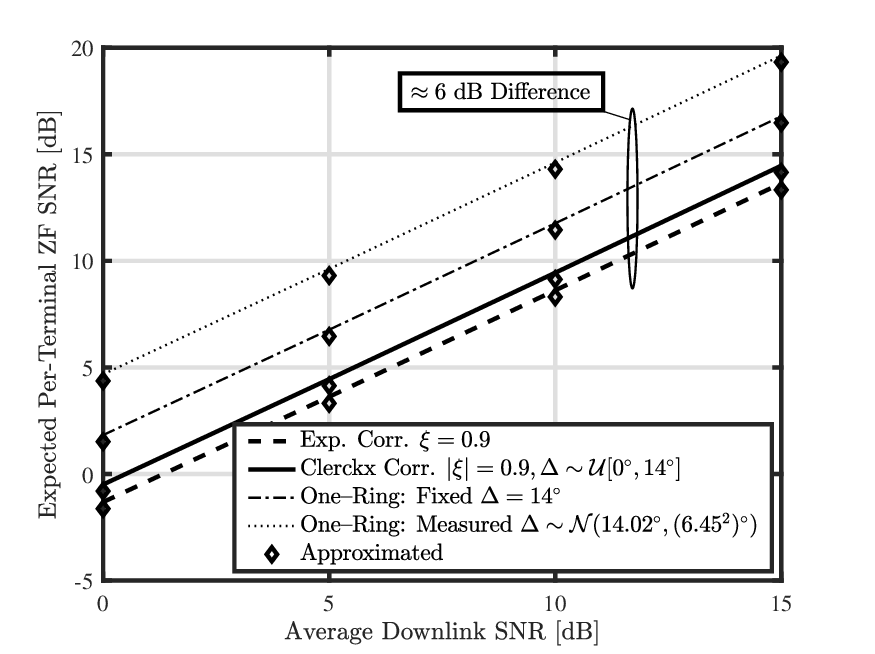}
\vspace{-6pt}
\caption{Expected ZF SNR vs. average SNR 
with $M=64$ and $L=6$.}
\label{expZFSNRvsSNR}
\vspace{-5pt}
\end{figure}

\begin{figure}[!] 
\hspace{-15pt}
\centering
\includegraphics[width=8.29cm]{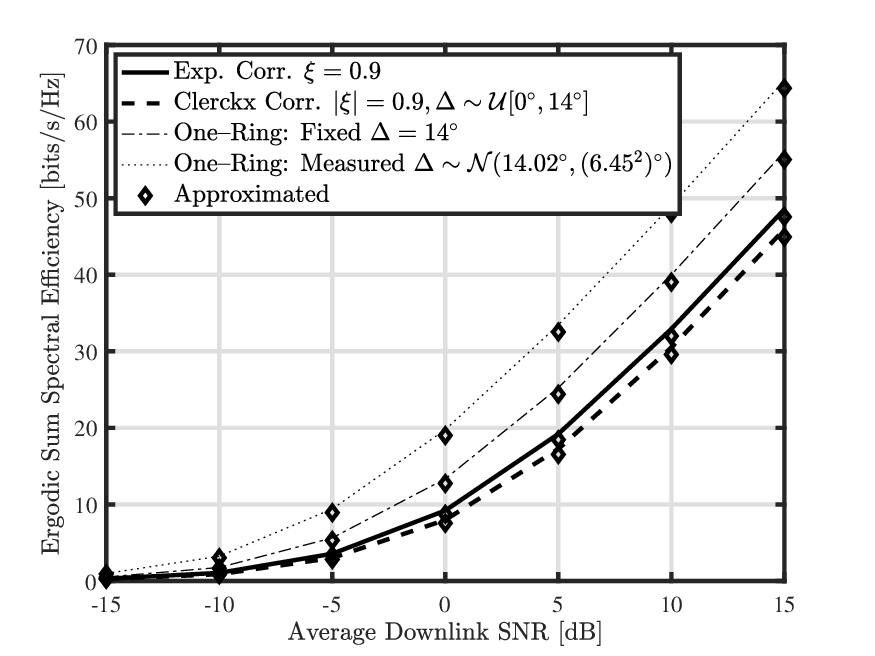}
\vspace{-4pt}
\caption{ZF Ergodic sum spectral efficiency vs. average SNR 
with $M=128$ and $L=10$.}
\label{ergSEvsSNR}
\vspace{-4pt}
\end{figure}

Figure~\ref{expZFSNRvsSNR} demonstrates the expected ZF SNR with changes in the  
average operating SNRs. In addition to the cases for exponential  
and Clerckx correlations, performance with the one--ring model is also evaluated. 
We can see that even with a \emph{fixed} angular spread, the one--ring model 
predicts higher ZF SNRs in comparison to the Clerckx model. This is because 
\emph{both} the magnitude and phase of the correlation matrices 
are variable across each terminal in the case of the one--ring model. 
\emph{Further to this, when evaluating the 
expected ZF SNRs with the measured angular parameters, 
a further 3 dB increase in the ZF SNR is seen across all 
SNRs. This is attributed to the increased diversity brought by the variations in 
the angular spreads (Gaussian random variables)}. 
The proposed approximations are seen to remain tight across all the considered 
models, and SNR values. 
Keeping all other parameters constant, Fig.~\ref{ergSEvsSNR} depicts 
the ZF ergodic sum spectral efficiency as a function of the operating SNR 
with $M=128$ and $L=10$. 
While similar trends to Fig.~\ref{expZFSNRvsSNR} 
can be observed, it is notable that the remarkably simple approximations 
remain tight across a wider range of system dimensions and 
operational SNRs.

\vspace{-2pt}
\section{Conclusions}
\label{Conclusions}
\vspace{-2pt}
The paper presents closed--form approximations to the ZF
expected SNR and ergodic sum spectral efficiency of a 
MU--MIMO system. With unequally correlated 
Rayleigh fading, our analysis is robust to 
various physical and non--physical channel 
correlation models, as well as average downlink SNR. 
More physically motivated models, such as one--ring 
correlation, consider unequal magnitudes and phases 
in the correlation matrices for each 
terminal, and tend to estimate higher MU--MIMO 
performance. Data from the recent 2.53 GHz
UMa propagation measurements was extracted to 
accurately parameterize correlation models in order to 
characterize their impact on MU--MIMO performance. 
Such an evaluation emphasizes the fact that 
the performance of a MU--MIMO system is ultimately 
governed by the correlation model, and its parameters 
in use.

\bibliographystyle{IEEEtran}

\end{document}